\documentclass[twocolumn,preprintnumbers,amsmath,amssymb,floats]{revtex4}
\usepackage{graphicx}
\usepackage{dcolumn}
\usepackage{bm}

\def\be{\begin{equation}}
\def\ee{\end{equation}}
\def\bea{\begin{eqnarray}}
\def\eea{\end{eqnarray}}

\def\half{{1\over 2}}

\begin{document}
\title{Coherence in the Two Kondo Impurity Problem}
\author{ Lijun Zhu and C. M. Varma}
\affiliation{Department of Physics and Astronomy, University of
California,
  Riverside, CA 92521}
\begin{abstract}
We show through a perturbative and an exact calculation using
Wilson's renormalization methods that in the problem of two
interacting Kondo impurities, on-site potential scattering generates
a quantum-tunneling between the two impurities through a  marginally
relevant operator. The magnitude of this tunneling $V_{12}$ depends
on the spin-correlation between the two impurities. For exchange
interactions between the moments comparable to the Kondo energy
$T_K$, $V_{12}$ is typically much larger than $T_K$. This implies
that the heavy-fermion mass in this interesting range is determined
by $V_{12}$ rather than $T_K$. The importance of these results for
experiments in coupled Kondo dots is also pointed out.
\end{abstract}
 \maketitle

The two Kondo impurity problem has been of interest because it
presents the competition between mutual interactions of local
moments and the Kondo renormalization of the moments to singlet
states in a soluble way. An aspect of the problem which has received
no attention hitherto is the the low-energy transfer Hamiltonian of
quasi-particles between the two impurity sites.  In the
quasi-particle problem for a lattice, this sets the scale for the
heavy-fermion bandwidth. The issue is also directly posed in a
recent ingenious realization of the two Kondo impurity problem using
coupled quantum dots \cite{Craig04}.

The two Kondo impurity problem was solved by numerical
renormalization group method (NRG) for a particle-hole symmetric
model obtained by ignoring potential scattering \cite{JonesV}, and
was subsequently studied by conformal field theory \cite{AffleckL},
bosonization\cite{sire} and other methods\cite{others}. In the
single Kondo impurity problem, potential scattering merely shifts
the Kondo resonance with respect to the chemical potential in
accordance with Friedel sum-rule. The sum of the phase shifts in the
even and the odd parity channels for the two Kondo impurity problem
is fixed by the Friedel sum-rule to be $\pi$. Without potential
scattering the phase shift in both the even and the odd parity
channels is $\pi/2$ when the fixed point is the Kondo fixed point.
Consequently, there is no splitting of the even-odd resonances or
equivalently, no {\it direct} hopping of the quasi-particles between
the two impurity sites. In this case there is a quantum critical
point (QCP) separating the Kondo singlet regime of the impurities
from the inter-impurity singlet regime \cite{JonesV}.

Affleck and Ludwig \cite{AffleckL} pointed out that the QCP is
changed to a crossover by the potential scattering only when the
exchange scattering at the local moments transfers conduction
electron from one impurity to the other. In this paper, we first
explain the Affleck-Ludwig result by a simple perturbative
calculation. We then obtain the new result that in such a case an
even-odd splitting of the resonances is generated which depends on
the spin correlations. We then present \cite{zarand} exact results
of a Wilson's NRG calculation to reinforce and quantify these
conclusions by deriving the low energy Hamiltonian. These also serve
to give the crossover scale of the QCP. The relation of the
coherence scale to the bare parameters of the problem and to the
generated spin-correlations is explicitly calculated. These results
are of fundamental importance to the problem of the heavy-fermion
state.

The interactions between the free electrons and the local moments in
the two Kondo impurity problem are \cite{JonesV, AffleckL}
\bea
H_{\text{imp}}&=& {J_+ \over 2} \left(  \psi^{\dag}_{1} {\vec
\sigma} \psi_{1} + \psi^{\dag}_{2} {\vec \sigma}
\psi_{2}\right) \cdot ({\bf S}_1 + {\bf S}_2) \nonumber \\
&+& {J_m \over 2} \left( \psi^{\dag}_{1} {\vec \sigma} \psi_{1} -
\psi^{\dag}_{2} {\vec \sigma} \psi_{2}\right) \cdot
({\bf S}_1 - {\bf S}_2) \nonumber \\
&+& { J_- \over 2} \left( \psi^{\dag}_{1} {\vec \sigma} \psi_{2} +
\psi^{\dag}_{2} {\vec \sigma} \psi_{1} \right) \cdot ({\bf S}_1 +
{\bf S}_2) \nonumber \\
&+& K {\bf S}_1 \cdot {\bf S}_2 + \sum_{\sigma} V \left(
\psi^\dag_{1\sigma}\psi_{1\sigma} +\psi^\dag_{2\sigma}\psi_{2\sigma}
\right).
\label{eq:H-2KI}
\eea
Here ${\bf S}_1$ and ${\bf S}_2$ represent two $S=1/2$ Kondo
impurities sitting at two lattice sites. $\psi_{1(2)}=\sum_{\bf k}
\psi_{1(2),k}$; $\psi_{1(2),k}$ are annihilation operators for
conduction electrons in spherical waves about the impurity $1$ and
$2$ respectively. ${\vec \sigma}$ are Pauli matrices.  $J$'s are the
exchange coupling between the local moments and the conduction
electrons. $V$ gives the potential scattering; the maximum magnitude
of $V$ deduced from the Anderson Hamiltonian is
$|J|/4$\cite{footnote}. It is convenient to choose certain $J_m$ so
that the generated RKKY interaction is 0 while introducing such an
interaction $K$ explicitly\cite{JonesV, footnote2}. For the NRG
calculation, the representation in terms of even ($e$) and odd ($o$)
parities of the conduction electron states is more convenient. The
latter are obtained by $\psi_{e(o),k} =(\psi_{1,k}\pm
\psi_{2,k})/\sqrt{2}$ while $J_{e,o}=J_+\pm J_-$.

\begin{figure}[tbh]
\centering
\includegraphics[width=0.6\columnwidth]{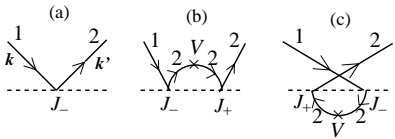}
\caption{Vertices diagrams for the two Kondo impurity problem. (a)
shows a bare vertex of $J_-$ while (b) and (c) show leading
corrections to $J_-$ in particle-particle and particle-hole
channels, respectively, which contribute a potential scattering
$V_{12} \psi^{\dag}_2 \psi_1$. Solid and dashed lines represent the
propagators of the conduction electrons and the two impurities with
a total spin ${\bf S}={\bf S}_1+{\bf S}_2$, respectively. }
 \label{fig:vertex}
\end{figure}

Note that in Eq.(\ref{eq:H-2KI}) the only connection between the
impurity states is through $J_-$.  (The direct hopping of the
impurity orbitals from one impurity site to the other is much much
smaller than $V$, $J's$ or the single-impurity Kondo temperature
$T_K$ for rare-earth or actinide moments.) This alone (i.e. with
$V=0$) is insufficient for quasi-particle hopping between sites $1$
and $2$. This can be motivated through a perturbative calculation by
the diagrams in Fig.(\ref{fig:vertex}).

The vertex can be in general separated as a vector and a scalar
part: $\Gamma_{12}= \Gamma_{S,12}\delta_{\alpha\alpha'} +
\Gamma_{V,12}{\vec \sigma}_{\alpha\alpha'} \cdot {\bf S}$. Here
${\bf S}={\bf S}_1+{\bf S}_2$. The scalar part, $\Gamma_{S, 12}$, is
an effective potential scattering electrons between sites $1$ and
$2$. First we consider the limit when two impurities are locked into
either a singlet or a triplet states, $S=0$, or $1$. The bare vertex
[cf. Fig. \ref{fig:vertex}(a)] is $\Gamma^{(0)}_{V,12}=J_-$. This is
corrected by Fig. \ref{fig:vertex}(b) and Fig. \ref{fig:vertex}(c)
in next leading order. They exactly cancel if the intermediate
particle-line in (b) and hole-line in (c) have the same spectra,
i.e. at $V=0$, or particle-hole symmetry. For $V \ne 0$, the
generated vertex (including similar diagrams with $\psi_{1k}$ as the
intermediate state) is:
\be
\Gamma_{S,12} = \rho^2 {S(S+1)\over 2}J_+ J_- \left(cV\right),
\ee
where $\rho$ is the density of state of conduction electrons at the
chemical potential, $S$ is the total spin of the two impurities and
$c$ is a dimensionless constant of $O(1)$. Such a term in two Kondo
impurity problem generates the new effect, a splitting between even
and odd resonances,
\be
V_{12} \psi_1^{\dag}\psi_2 + h.c. = V_{12}(\psi_e^{\dag}\psi_e
-\psi_o^{\dag}\psi_o). \label{eq:coh-scale}
\ee
$V_{12}$ also depends on the spin-states of the two impurities; this
is seen from the fact that $V_{12}=V_e-V_o$ while the effective
potential scattering including Fig. \ref{fig:vertex}(b-c) is
 ${V}_e \psi^\dag_e\psi_e + {V}_o \psi^\dag_o\psi_o$
\be
{V}_{e,o} = V - {S(S+1)\over 4} c V \rho^2(J_+^2 + J_-^2 \pm 2
J_+J_-)
\ee
When $K\to-\infty$, $S=1$; then
\be
{V}_{12} \propto V (\rho J_-)(\rho J_+).
\ee
However, when $K\to\infty$, $S=0$, the contribution to the splitting
due to $V$ vanishes. In between these limits, there are mixing terms
between the triplet and singlet states of two impurities, so that
the results depend on the spin-correlation $\langle {\bf S}_1 \cdot
{\bf S}_2\rangle$. For the general case and because at least one of
the two vertices in Fig.(\ref{fig:vertex}), $J_+$, has singular
renormalizations, complete answers can only be obtained by exact
methods, for example, Wilson's NRG method\cite{Wilson}.

In NRG, the conduction electron band is discretized on a logarithmic
grid into a semi-infinite chain with only nearest-neighbor hopping
terms (here we have two chains corresponding to even and odd parity
degrees of freedom), while the model can be solved recursively by a
sequence of $H_N$'s.
\be
H_{N+1}=\Lambda^{1/2}H_N +\sum_{p=e,o;\sigma}
[f^\dag_{pN\sigma}f_{p(N+1)\sigma}+\text{H.c.}],
\label{eq:H-iter}
\ee
where $\Lambda$ is the discretization energy scale. The impurities
only interact with the first site ($f_{p0\sigma}$) through the
Hamiltonian $H_0$. From the change of spectra with iteration, the
fixed points and the dimensions of operators about them are
determined.

\begin{figure}[tbh]
\centering
\includegraphics[width=0.8\columnwidth]{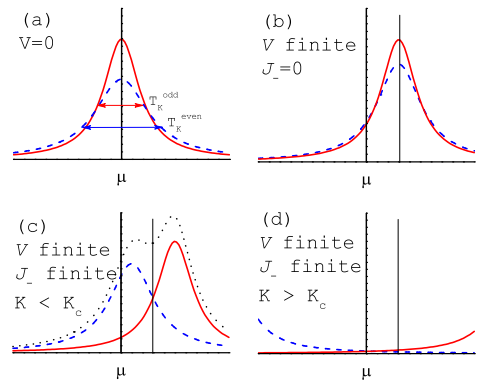}
\caption{Spectra for the two Kondo impurity model. The location of
the resonance with respect to the chemical potential is determined
from the fixed point Hamiltonian and the width from the leading
irrelevant operators. (a) shows the particle-hole symmetric case;
Kondo resonances of both even and odd parities appear at the
chemical potential. (b) for $J_-=0$ while finite potential
scattering $V$ breaks the particle-hole symmetry in each channel;
both resonances appear at the same energy but tuned away from the
chemical potential. (c) for both finite $V$ and $J_-$ while $K<K_c$;
the even and odd resonances split. For (d), when $K$ is bigger than
the crossover scale $K_c$, although the splitting increases, the
resonance peaks fall out of the band, reflecting the disappearance
of the Kondo effect.} \label{fig:spectrum}
\end{figure}

For $V=0$, two distinct spectra are identified for large iterations
with the tuning of $K$ \cite{JonesV}. When $K$ is large and
ferromagnetic, the spectra can be fitted with the effective
Hamiltonian
\be
H_N = H_N^* + H_{ir},
\ee
where the fixed-point Hamiltonians $H_N^*$ in even/odd iterations
are the same as those of free-electrons in odd/even iterations,
reflecting the $\pi/2$ phase shift in both parity channels and the
formation of Kondo resonances. Both resonances develop at the
chemical potential, as shown in Fig.(\ref{fig:spectrum}a). $H_{ir}$
contains terms such as $t_p f^{\dag}_{0p} f_{1p} +H.c.$ which vanish
at $N\to\infty$ as $\Lambda^{-(N-1)/2}$ reflecting the irrelevant
nature. ($t_p$ and other leading irrelevant parameters serve to
determine $T_K$). When $K$ is large and antiferromagnetic, the two
impurities form a spin-singlet; the eigenvalues of the actual
Hamiltonian do not show the even/odd interchange, reflecting 0 or
$\pi$ (for odd and even parity channels, respectively) asymptotic
phase shifts. Near the QCP at $K_c \approx 2.4 T_K^0$, singular
fermi-liquid behavior, studied in detail
earlier\cite{JonesV,AffleckL} ensues.

We find in accordance with the perturbative calculation above that
the fermi-liquid behavior on either side and the singular
fermi-liquid behavior near the QCP continues for  $V \ne 0$, $J_-
=0$ but with equal additional phase shift in both parities so that
the asymptotic eigenvalues are given by
\be
\sum_{p}V^*_p f^{\dag}_{0p}f_{0p},
\ee
with $V_e^*=V_o^*$. This has a scaling dimension 0 and appears as a
marginally relevant operator at the two strong-coupling fixed points
as well as at the QCP. The conservation of {\it axial charge}
\cite{JonesV, AffleckL} continues to give relation between the
coefficients for the irrelevant operators and they diverge in the
same manner as at $V=0$ near the QCP. When $J_-=0$, the two
impurities become effectively two non-interacting Kondo scatters.
For each channel, a potential scattering induced phase shift
$\delta_v = -\tan^{-1}(\pi\rho_0 V)$ appears in addition to the
$\pi/2$ phase shift, as in single Kondo impurity
problem\cite{Cragg78}. Here, $\delta_e=\delta_o$, two Kondo peaks in
local density of states coincide with each other while they are
shifted away from the chemical potential [see
Fig.(\ref{fig:spectrum}b)].

\begin{figure}[tbh]
\centering
\includegraphics[width=1.0\columnwidth]{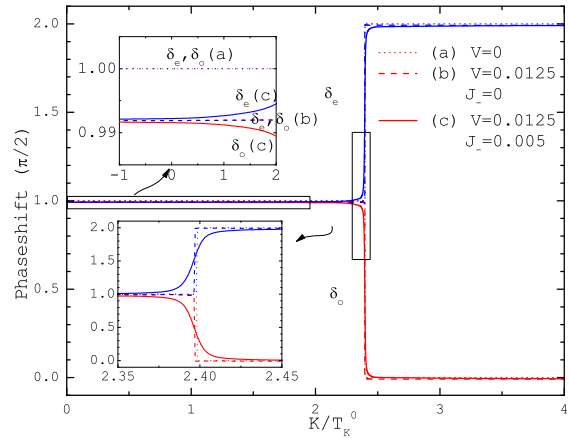}
\caption{Phase shifts as functions of $K$. (a)-(c) show the cases
when $V=0$, $J_-=0.005$; $V=0.0125$, $J_-=0$; and $V=0.0125$,
$J_-=0.005$, respectively. $J_+$ is chosen as $0.125$.}
\label{fig:phaseK}
\end{figure}

It is natural to represent the eigenvalues in the even and odd
parity channels in terms of the phase-shifts $\delta_e$ and
$\delta_o$ in these channels, which can obtained from the
eigenvalues of the fixed-point Hamiltonian, denoted as
$\omega^*_{pj}$ for $j$-th excitation level with $p$-parity. For
free electrons(with vanishing $V$ and $J$'s), the eigenvalues
($\eta^*_j$) can be easily obtained numerically for a given
$\Lambda$. Their values at $N\to\infty$ limit are approximately
\bea
\text{$N$ odd:} && \eta^*_j = \text{sgn}(j) \Lambda^{|j|-1},
j=\pm1,\ldots,\pm \half(N+1), \nonumber \\
\text{$N$ even:} && {\hat\eta}^*_j = \text{sgn}(j)
\Lambda^{|j|-1/2}, j=0,\pm1,\ldots,\pm \half N,
\eea
for $1\ll|j|\ll N/2$. With interactions, the
eigenvalues($\omega^*_j$) can be associated with $\eta^*_j$ by a
phaseshift factor. Quite generally,
\be
\omega^*_{pj} = \Lambda^{-\text{sgn}(\eta^*_{pj})\delta_p/\pi}
\eta_{pj}^*.
\label{eq:phase-def}
\ee
For instance, at the strong-coupling Kondo fixed point, ${\hat
\omega}^*_j$ at even iterations ($\Lambda^{j-1/2} \Lambda^{-1/2}$)
are the same as $\eta^*_j$ in odd iterations, corresponding to a
phase shift $\delta_p=\pi/2$. One can then calculate the phase
shifts from the numerical data following Eq.(\ref{eq:phase-def}).
The principal results of this paper is the evaluation of $\delta_e$,
$\delta_o$ and its dependence on $K/T_K^0$ from which the variation
of the coherence scale is determined. These results are displayed in
Fig. (\ref{fig:phaseK}).

In the particle-hole symmetric case ($V=0$), $\delta_e=\delta_o=
\pi/2$ in the Kondo-resonance fixed point, but $\delta_e=\pi$ and
$\delta_o=0$ in the inter-impurity spin-singlet fixed point. The
latter merely reflects the fact that in the inter-impurity
spin-singlet state, the even parity channel is fully occupied and
the odd parity channel is completely empty. An abrupt change of the
phase shift happens at the critical point, $K_c\approx 2.4 T_K^0$,
where $T_K^0 \approx 1.4\times 10^{-4}D$, is the single-impurity
Kondo temperature (for $J/D=0.125$). $J_{-}$ is then irrelevant.
When $V\neq0$ but $J_{-}=0$, $\delta_e$ and $\delta_o$ of two phases
are shifted equally from $\pi/2$ and $\pi$ or $0$, respectively.
They still behave as two independent Kondo impurities; each gains
additional $\delta_v$ from the potential scattering term. With both
finite $V$ and $J_{-}$, one can observe two features. 1) the phase
shift changes smoothly at a critical value for $K_c(V)$, which
smears out the critical point. $K_c$ is a function of $V$ since the
potential scattering term renormalizes the Kondo coupling (see
perturbative calculation above) thus changing the Kondo temperature.
2) even and odd phase shifts are different while their summation
remains $\delta_e+\delta_o=\pi+2\delta_v$, where $\delta_v$ can be
associated with its counterpart in one Kondo impurity problem. In
this case, although the axial charge in each channel is not
separately conserved, their summation is. $J_-$ is a relevant
operator removing the critical point if particle-hole symmetry is
absent, $V \ne 0$.

\begin{figure}[tbh]
\centering
\includegraphics[width=0.9\columnwidth]{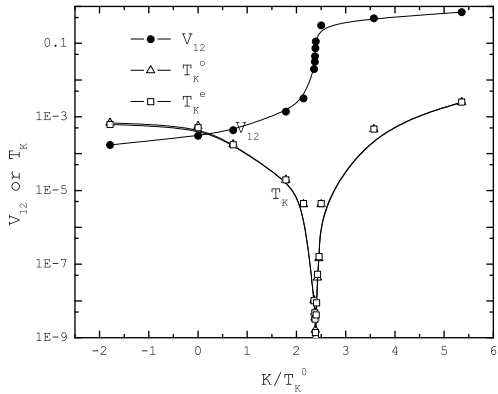}
\caption{The dependence of $V_{12}$ and the Kondo temperature $T_K$
(both in unit of the bandwidth $W$) on the RKKY coupling constant
$K$. Other parameters are fixed as $J_+=0.125$, $J_-=0.005$ and
$V=0.0125$.}
\label{fig:deltaK}
\end{figure}

\begin{figure}[tbh]
\centering
\includegraphics[width=1.0\columnwidth]{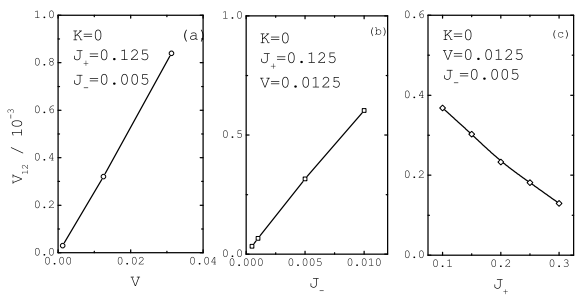}
\caption{The dependence of the coherence scale $V_{12}$ on various
parameters: a) $V$; b) $J_-$; c) $J_+$ when $K=0$.}
\label{fig:delta}
\end{figure}

Given the phase shifts the potential scattering parameter in the
asymptotic Hamiltonian which is a marginal operator are determined
through $\delta_{e,o}= -\tan^{-1}(\pi\rho V_{e,o}^*)$,
\be
(V_e^*-V_o^*)(f^{\dag}_{e0}f_{e0}-f^{\dag}_{o0}f_{o0}),
\ee
which is the coherence operator in effective Hamiltonian
\be
H^*_{coh}= V_{12} (\psi^\dag_1\psi_2+H.c.), \label{eq:H-coh}
\ee
where $V_{12}=D (V_e^*-V_o^*)(1+\Lambda^{-1})/2$, i.e, $V_{12}$ in
Eq.(\ref{eq:coh-scale}) for $\Lambda =1$.

The dependence on the coherence scale $V_{12}$ on various parameters
is shown in Figs.(\ref{fig:deltaK},\ref{fig:delta}).  $V_{12}$ shows
a linear relation with $V$ and $J_-$ for a whole range of $K$, as
shown in Fig.(\ref{fig:delta}a-b). This is in agreement with the
perturbation results in $K\to-\infty$ limit, while as expected its
dependence on $J_+$ for a general $K$ is not. The critical point
becomes a crossover, as the degeneracy of the singlet and triplet
parts of the Hamiltonian is lifted. Close to the crossover, although
$V_{12}$ increases uniformly, the Kondo resonances themselves
disappear [cf. Fig.(\ref{fig:deltaK})].

Our most important new physical results come from the magnitude of
$V_{12}$ and its dependence on $K/T_K^0$.  In the particle-hole
symmetric case, such a term vanishes: the quasi-particle does not
hop {\it directly} between the two sites. (The analog of this effect
in the periodic Anderson lattice model, is that the chemical
potential falls within the hybridization gap for the particle-hole
symmetric case and the system is an insulator \cite{Pruschke}). A
finite $V_{12}$ has important implications for the heavy-fermion
lattice. Heavy-fermion bands have been derived in a variety of ways,
all of which have the same physics: constrained hybridization of a
periodic array of local orbitals with a wide conduction band with a
matrix element of the order of the single Kondo impurity resonance
width, i.e., the Kondo temperature. The heavy-fermion bandwidth in
such calculations is therefore of $O(T_K)$. However even for $T_K
\to 0$, $V_{12}$ gives the order of magnitude of the splitting of
the heavy fermion band at the zone-center and the zone-boundary.
Therefore, if the lattice problem is solved by a self-consistent two
Kondo impurity problem with a cluster generalization of dynamical
mean-field theory, the effective bandwidth is then to be determined
including both the parameter $T_K$ as well as $V_{12}$. It is
therefore important to consider the relative magnitude of these two
parameters. The interesting range for heavy-fermion compounds is for
antiferromagnetic coupling $K/T_K^0$ of $O(1)$. In this range for
the typical parameters shown in Fig.(\ref{fig:deltaK}), $V_{12}$ is
more than an order of magnitude larger than $T_K$.
The conclusion then is that the effect of $V_{12}$ dominates in the
low energy  mass and  thermodynamic and  transport properties. It is
interesting that this term already includes renormalizations due to
spin-correlations, which are further augmented by Landau-parameters.

In the regime of large $K/T_K^0$, the $T_K$ given in
Fig.(\ref{fig:deltaK}) reflects only that the spectral weight at the
chemical potential is only of $O(T_K/V^2_{12})$. This reflects the
disappearance of the Kondo resonance for such parameters and its
replacement by the spectra schematically illustrated in Fig.
(\ref{fig:spectrum}d).

Recently, double quantum-dot systems have been fabricated to
simulate the two Kondo impurity model \cite{Craig04}. When the two
dots are allowed to interact, two split Kondo resonances appear in
the tunneling spectra. The spectrum in Fig.(3a) of
Ref.\cite{Craig04} is similar to our Fig.(\ref{fig:spectrum}c), and
the two observed peaks can be identified as the even and odd parity
resonances of the Kondo dots.

This work was partially supported by the LANL--UCR CARE
Collaborative Research Programs.


\begin{references}
\bibitem{footnote2}
We pick $K>0$ to be antiferromagnetic while $K<0$ to be
ferromagnetic, while in Ref. \cite{JonesV} another notation $I_0$ is
used with opposite signs.
\end{references}
\end{document}